# Solar dynamo and geomagnetic activity

Katya Georgieva[1], Boian Kirov[2]

1 - Solar-Terrestrial Influences Laboratory, Bulgarian Academy of Sciences, e-mail: kgeorg@bas.bg
2 - Solar-Terrestrial Influences Laboratory, Bulgarian Academy of Sciences, e-mail: bkirov@space.bas.bg

## Abstract

The correlation between geomagnetic activity and the sunspot number in the 11-year solar cycle exhibits long-term variations due to the varying time lag between the sunspot-related and non-sunspot related geomagnetic activity, and the varying relative amplitude of the respective geomagnetic activity peaks. As the sunspot-related and non-sunspot related geomagnetic activity are caused by different solar agents, related to the solar toroidal and poloidal fields, respectively, we use their variations to derive the parameters of the solar dynamo transforming the poloidal field into toroidal field and back. We find that in the last 12 cycles the solar surface meridional circulation varied between 5 and 20 m/s (averaged over latitude and over the sunspot cycle), the deep circulation varied between 2.5 and 5.5 m/s, and the diffusivity in the whole of the convection zone was ~$10^8$ m²/s. In the last 12 cycles solar dynamo has been operating in moderately diffusion dominated regime in the bulk of the convection zone. This means that a part of the poloidal field generated at the surface is advected by the meridional circulation all the way to the poles, down to the tachocline and equatorward to sunspot latitudes, while another part is diffused directly to the tachocline at midlatitudes, "short-circuiting" the meridional circulation. The sunspot maximum is the superposition of the two surges of toroidal field generated by these two parts of the poloidal field, which is the explanation of the double peaks and the Gnevyshev gap in sunspot maximum. Near the tachocline, dynamo has been operating in diffusion dominated regime in which diffusion is more important than advection, so with increasing speed of the deep circulation the time for diffusive decay of the poloidal field decreases, and more toroidal field is generated leading to a higher sunspot maximum. During the Maunder minimum the dynamo was operating in advection dominated regime near the tachocline, with the transition from diffusion dominated to advection dominated regime caused by a sharp drop in the surface meridional circulation which is in general the most important factor modulating the amplitude of the sunspot cycle.

## 1. Introduction

Though solar activity is the source of geomagnetic disturbances (see e.g. Cliver, 1994 for a review), the correlation in the 11-year solar cycle between solar activity expressed by the sunspot number and geomagnetic activity measured by the aa-index (Mayaud, 1972) has been decreasing since 1868, the beginning of the aa-index record. Kishcha et al. (1999) suggested that the long-term variations in the sunspot-geomagnetic correlation may result from the quasi-periodic fluctuations of the time lag of the geomagnetic indices relative to sunspot numbers. Echer et al. (2004) confirmed that the correlation between geomagnetic and sunspot activity in the 11-year solar cycle has decreased since the end of the 19th century, and the lag between them has increased. As a probable cause of the correlation decrease they pointed at the aa-index dual peak structure: the second aa-index peak related to the high-speed solar wind streams seems to have increased relative to the first one related to sunspot (coronal mass ejections) activity. In an earlier paper we (Georgieva and Kirov, 2006) showed that the correlation between sunspot and geomagnetic activity has decreased not only in the 11-year cycle but also on long-term time-scale.

Sunspot-related solar activity whose manifestations are the solar coronal mass ejections, is the cause of the strongest individual geomagnetic storms in all phases of the solar cycle (Webb, 2002). However, such strong storms are short-lasting and relatively rare, and have little impact to the average yearly geomagnetic activity, except around sunspot maximum. Non-sunspot-related solar activity whose manifestations are the high-speed solar wind streams from long-lived solar coronal holes, causes geomagnetic storms which are as a rule weaker, but long-lasting and recurrent, so they provide the main contribution to the average yearly geomagnetic activity. As a result, the variations in geomagnetic activity are better correlated to the variations in non-sunspot-related solar activity than in the sunspot-related solar activity (Richardson and Cane, 2002). Therefore, the variations in geomagnetic activity can be used as a proxy for the variations in non-sunspot-related solar activity. To check this, in the next section we implement the method first proposed by Feynman (1982) to divide the long-term solar activity variations into sunspot-related and non-sunspot-related. In Section 3, after a brief description of the Babcock-Leighton flux-transport solar dynamo mechanism, we introduce our method to derive the long-term variations in solar surface and deep meridional circulations and in the turbulent diffusivity in the bulk of the convection zone based on geomagnetic and sunspot data, and relate them to the long-term variations in solar activity. In Section 4 we use the derived variations in the solar meridional circulation to test the validity of the flux-transport dynamo mechanism, and try to identify the factors determining the long-term variations in solar activity. Finally, in Section 5 we summarize and discuss our results.

## 2. Long-term variations in sunspot-related and non-sunspot-related solar activity

Fig.1a presents the geomagnetic aa-index and the international sunspot number, monthly averages, smoothed with the well known Gleissberg filter (Gleissberg, 1944). In Fig.1b their sunspot cycle averages are shown. It can be seen that the aa-index and sunspot number variations are not identical, neither in the 11-year solar cycle nor in the long-term trend. Moreover, as obvious from Fig.1b, the changing correlation between them is due not only to the changing time lag between their peaks in the 11-year sunspot cycle but also due to their changing relative amplitudes.

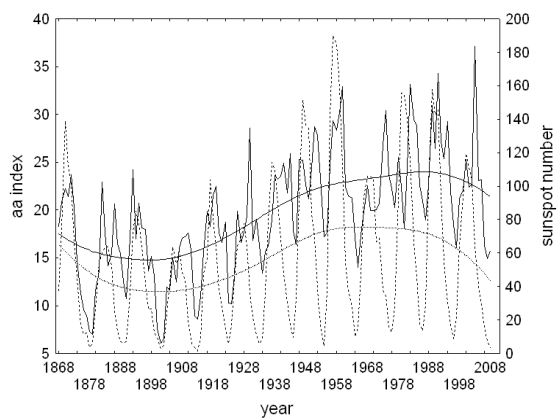

Fig.1a: aa-index of geomagnetic activity (solid line) and sunspot number (dotted line), monthly values, 13-point moving averages. The bold solid and dotted lines are the least square fits to aa-index and sunspot number, respectively.

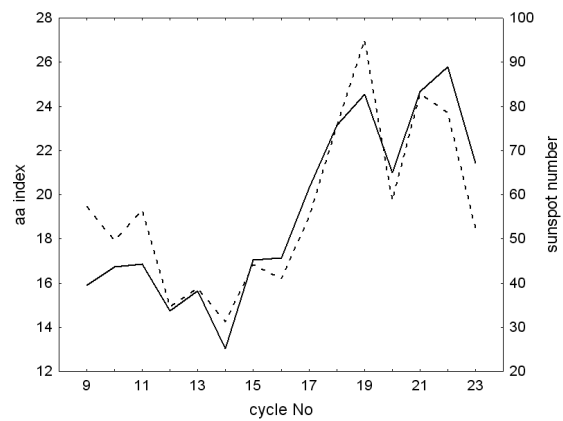

Fig.1b: Long-term variations of the aa-index of geomagnetic activity (solid line) and sunspot number (broken line), sunspot cycle averages.

As mentioned above, geomagnetic activity can be caused by two types of solar drivers, manifestations of two types of solar activity: sporadic or sunspot-related, and recurrent or non-sunspot-related. If we can separate the contribution of these two types of solar drivers to geomagnetic activity, we can estimate the relative intensity of the two types of solar activity. Feynman (1982) noticed that if we plot geomagnetic aa-index as a function of the sunspot number, all points lie above a line – that is, for every level of sunspot activity there is a minimum value of geomagnetic activity (Fig.2).

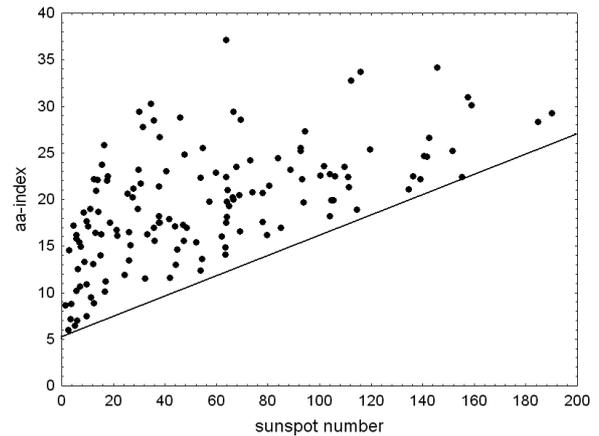

Fig.2: Dependence of the aa-index of geomagnetic activity on the sunspot number.

The equation of this minimum line indicates what part of the geomagnetic activity ($aa_T$ - the sporadic geomagnetic activity) is caused by sunspot-related solar activity. An updated equation based on a longer time-series is derived by Ruzmaikin and Feynman (2001):

$$aa_T = 0.07 * R + 5.17$$

where R is the sunspot number. What is left ($aa_P = aa - aa_T$ - the recurrent geomagnetic activity) is caused by non-sunspot-related solar activity.

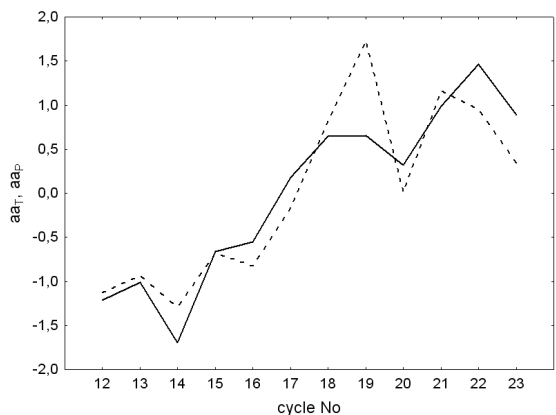

Fig.3: Long-term variations of non-sunspot-related (solid line) and sunspot-related (broken line) geomagnetic activity, sunspot cycle averages.

Fig.3 presents the long-term variations (sunspot cycle averages) of $aa_T$ and $aa_P$. As the exact values of $aa_T$ and $aa_P$ strongly depend on the correct choice of parameters in the above equation, they are presented in units of standard deviations. It can be seen that the relative variations in $aa_T$ and $aa_P$ are not identical, and are very similar to the relative variations of the sunspot number and aa-index.

Now, the following questions arise: what is the cause of the long-term geomagnetic and sunspot activity variations, and what is the cause of the difference in the long-term variations of sunspot-related and non-sunspot-related solar activity. To address these questions, we turn to the solar dynamo mechanism which drives the solar activity: transforming the poloidal (non-sunspot-related) solar magnetic field into toroidal (sunspot-related) solar magnetic field, and of this toroidal field back into poloidal field with the opposite magnetic polarity.

## 3. The solar dynamo mechanism and the importance of the meridional circulation

### 3.1. The flux-transport solar dynamo mechanism

The rotation in the upper part of the Sun, down to about 0.7 $R_s$ (the so-called convection zone) is differential: fastest at the equator and decreasing toward the poles, while the radiative zone below (between 0.7 $R_s$ and the core spreading up to 0.2 $R_s$) rotates rigidly. At the base of the convection zone, the differential rotation stretches the "poloidal" (north-south) field in azimuthal direction giving rise to the "toroidal" (east-west) component of the field - a process generally known as "Ω-effect" (Parker, 1955). The buoyant magnetic field tubes rise up, piercing the surface at two spots (sunspots) with opposite magnetic polarities. Due to the Coriolis force, the bipolar pair of spots is tilted with respect to the meridional plane with the leading (in the direction of solar rotation) spot at lower heliolatitude than the trailing spot. In each solar hemisphere, the leading spot has the polarity of the respective pole, and the trailing spot has the opposite polarity.

This part of the dynamo - the transformation of the poloidal field into toroidal field - is believed to be clear, while the physical mechanism responsible for the regeneration of the poloidal component of the solar magnetic field from the toroidal component (the so called α-effect) has not yet been identified with confidence (Charbonneau, 2005). Different classes of mechanisms have been developed. Recently the most promising seems to be the one based on the idea first proposed by Babkock (1961) and mathematically developed by Leighton (1969): late in the sunspot cycle, the leading spots diffuse across the equator where their flux is canceled by the opposite polarity flux of the leading spots in the other hemisphere. The flux of the trailing spots and of the remaining sunspot pairs is carried toward the poles where it first cancels the flux of the previous solar cycle and then accumulates to form the poloidal field of the next cycle with polarity opposite to the one in the preceding cycle. Wang et al. (1991) suggested that this mechanism includes a meridional circulation with a surface flow toward the poles where the poloidal flux accumulates, sinks to the base of the solar convection zone, and is carried by the counterflow there back to low latitudes to be transformed into toroidal flux and to emerge as the sunspots of the nest solar cycle.

This so-called flux-transport dynamo mechanism is directly observed, both the forming of unipolar magnetic regions from decaying sunspot pairs, and the meridional circulation carrying them to the poles. The near-surface poleward flux has been confirmed from helioseismology (Hathaway, 1996, and the references therein; Makarov et al., 2001; Zhao and Kosovichev, 2004; Gonzalez Hernandez et al., 2006), magnetic butterfly diagram (Ivanov et al., 2002; Švanda et al., 2007), latitudinal drift of sunspots (Javaraiah and Ulrich, 2006). The deep counterflow has not been yet observed, but has been estimated by the equatorward drift of the sunspot occurrence latitudes (Hathaway et al., 2003; 2004). In what follows we assume that it is this flux-transport dynamo which operates in the Sun.

### 3.2. Role of the meridional circulation and the diffusion for the amplitude of the sunspot cycle

The long-term variations in the speed of both the surface and the deep meridional circulation should lead to long-term variations in solar activity. The surface circulation is important for the generation of the solar poloidal field of the new cycle from the toroidal field of the old cycle. According to Wang et al. (2002), the effect of a fast poleward flow at low latitudes is to weaken the polar field at the end of the cycle: If the surface circulation is faster, the leading polarity flux will not have enough time to diffuse across the equator and to cancel with its counterpart of the opposite hemisphere, so both trailing polarity flux and leading polarity flux will be carried poleward. They will first cancel each other so less uncanceled trailing polarity flux will be left to form the poloidal field of the new cycle. From this weaker poloidal field, a weaker toroidal field will be generated, and respectively the maximum sunspot number in the next cycle will be smaller.

If the diffusivity in the upper part of the convection zone involved in poleward motion is low, all of the generated poloidal field is advected by the meridional circulation all the way to the poles, down to the tachocline and equatorward to sunspot latitudes to be transformed into the toroidal field of the next cycle. Following the terminology of Hotta and Yokoyama (2010), this regime can be specified as "advection-dominated near the surface".

If the diffusivity in the upper part of the convection zone is high enough, a part of the poloidal field can diffuse directly towards the mid-latitude tachocline where the toroidal field of the next cycle is produced, and another part completes the full circle to the poles, down to the tachocline, and back to midlatitudes. The model experiments of Jiang et al. (2008) have shown that high enough diffusivity is $\eta_{surf}$ ~ $1-2.10^8$ m²/s, and not high enough is at least an order of magnitude lower. We can specify this regime as "moderately diffusion-dominated near the surface".

Hotta and Yokoyama (2010) have introduced a regime with even higher surface diffusivity in which all of the poloidal field generated at the surface is diffused down to the tachocline without being carried by the meridional flow to the poles. They call this regime "strongly diffusion-dominated near the surface", and show that the conditions for the dynamo to operate in this regime are $\eta_{surf}$ = $2 \div 9 \times 10^8$ m²/s and $\eta_{surf}/u_0 > 2 \times 10^7$ m, where $u_0$ is the maximum surface meridional flow speed.

The magnitude of the toroidal field generated from the poloidal field during its transport at the base of the solar convection zone depends on the interplay between the speed of the deep equatorward meridional circulation and the poloidal field diffusivity in the deeper part of the convection zone participating in the equatorward circulation (Yeates et al., 2008). In the diffusion-dominated near the tachocline regime, when the diffusivity is relatively high and the circulation is relatively slow, with increasing speed of the flow the time for diffusive decay of the poloidal field during its transport through the convection zone decreases, so more toroidal field is generated and respectively the number of sunspots in the following maximum is higher. In the advection-dominated near the tachocline regime, when the speed of the equatorward circulation is relatively high and the diffusivity there is relatively low, the diffusivity is not so important and with increasing speed of the flow the time to induct toroidal field in the tachocline decreases which leads to a weaker toroidal field and respectively a lower number of sunspots.

### 3.3. Deriving the long-term variations in the meridional circulation

The comparison of the variations in the speed of the deep and surface meridional circulation, the diffusivity, and the amplitude of the sunspot cycle is vital for verifying the solar dynamo theory and understanding the regime in which the dynamo operates. However, neither the speed of the deep equatorward circulation nor the magnetic diffusivity in the solar convection zone are known from observations. There are some indications for variations in the speed of the surface meridional circulation in the course of the sunspot cycle (Basu and Antia, 2003; Javaraiah and Ulrich, 2006; Javaraiah, 2010), however its long-term variations are not known.

We have proposed a method for estimating the long-term variations in the surface and deep meridional circulation from geomagnetic data (Georgieva and Kirov, 2007), based on the varying time lag between the maximum sunspot number in the 11-year solar activity cycle and the following geomagnetic activity maximum on the sunspot declining phase. Here we briefly explain this method and extend it to determine also the diffusivity in the bulk of the convection zone.

For our estimations we assume one circulation cell per hemisphere, and one full circulation circle per sunspot cycle. Numerical simulations of solar-like turbulent convection (Miesch et al. 2000) suggest multiple cells of varying lifetimes, but when averaged over 10 solar rotations or more, the mean flow pattern emerges consisting of a single large flow cell. As we are concerned with even longer time-scale (of the order of the sunspot cycle), this approximation is well justified. We should note here that all quantities which we derive are sunspot cycle averages over latitude and over time. Actually, as shown by Hathaway et al. (2003) and Javaraiah and Ulrich (2006) among others, the speed of the meridional circulation varies with both latitude and time, decreasing from middle to high and low latitudes, and from the beginning toward the end of the sunspot cycle.

### 3.3.1. Surface poleward circulation

The maximum geomagnetic activity during the sunspot declining phase is due to the interaction of the Earth's magnetosphere with high speed solar wind from big long-lived low latitude coronal holes and equatorward extensions of polar coronal holes. Low latitude coronal holes are formed from the remnants of the sunspot pairs. Just after sunspot maximum, these low latitude holes are small and narrow, and the Earth is only embedded for a short time in the high speed solar wind emanating from them. When the trailing polarity flux reaches the poles and forms the polar coronal holes, the low latitude holes begin attaching themselves to the polar holes and start growing, so the Earth is embedded in long-lasting wide streams of fast solar wind (Wang and Sheeley, 1990) leading to peak geomagnetic activities. Therefore, we can assume that the moment of the geomagnetic peak on the sunspot declining phase is the moment when the trailing polarity flux has reached the poles, and the time between the sunspot maximum and this geomagnetic peak which we denote by $t_{SG}$, is the time it takes the surface meridional flow to carry the flux from sunspot maximum latitudes (~15°) to the poles. In case the sunspot maximum is double peaked, for determining $t_{SG}$ we choose the higher peak. From $t_{SG}$ we can calculate $V_{surf}$ - the speed of the surface meridional circulation. With the solar radius equal to $6.96 \times 10^8$ m, these 75° on the solar surface are $9.11 \times 10^8$ m. If we divide this value by $t_{SG}$ expressed in seconds (= months x $2.6352 \times 10^6$), we get directly the speed of the surface circulation $V_{surf}$ in m/s. In the period 1868-2000 $V_{surf}$ varies between 5 and 20 m/s (Fig.4). $V_{surf}$ agrees remarkably well with results from helioseismology and magnetic butterfly diagrams which show latitude-dependent speed profile smoothly varying from 0 m/s at the equator to 20-25 m/s at midlatitudes to 0 m/s at the poles. Also shown in Fig.4 is the amplitude of the following sunspot cycle. As predicted by theory (Wang et al., 2002), $V_{surf}$ is anticorrelated with the amplitude of the following sunspot cycle (r=-0.7 with p=0.03): the slower the surface poleward meridional circulation, the higher the amplitude of the following sunspot maximum. Note that in Fig.4, $V_{surf}$ is shifted by one cycle relative to the sunspot maximum amplitude: that is, $V_{surf}$ in cycle n is compared to the amplitude of the sunspot maximum in cycle n+1. Besides, $V_{surf}$ is drawn with a reversed scale to better illustrate the correlation.

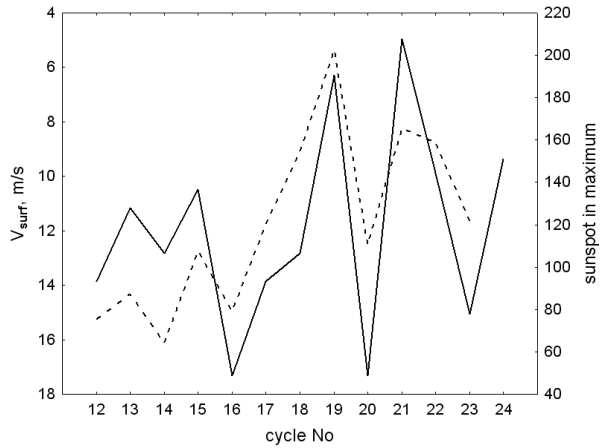

Fig.4: Amplitude of the sunspot maximum in consecutive sunspot cycles (broken line) and the speed of the surface meridional circulation preceding this sunspot maximum (solid line). Note the reversed scale of Vsurf.

### 3.3.2. Deep equatorward circulation

**A. Advection-dominated regime near the surface**

If the diffusivity in the upper part of the convection zone is low, the time between the geomagnetic activity maximum on the sunspot decline phase and next sunspot maximum ($t_{GS}$) is the time it takes the flux which has been carried to the polar region by the surface meridional circulation to sink there to the base of the convection zone, to be carried by the deep equatorward circulation to sunspot latitudes, and to emerge as the sunspots of the new cycle. Here again, as for calculating $V_{surf}$, in case of double peaked cycles $t_{GS}$ is the time between the geomagnetic activity peak on the sunspot declining phase and the highest following sunspot maximum. The time for the flow to emerge to the surface at sunspot latitudes is believed to be of the order of a month and we assume this value. The time for the flow to sink to the base of the convection zone in the polar region is not known. If the diffusivity is low, this time will be determined by the speed of the flow. We have calculated the speed of the deep meridional circulation in two cases: assuming the speed of the downward transport equal to the speed of either the surface or the deep circulation. The difference between the two results is 7 - 8 %. In what follows we use the values calculated with the speed of the downward transport equal to the speed of the deep meridional circulation. Therefore, the distance traversed by the deep circulation from the pole (90°) to sunspot maximum latitudes (15°) will be 75° at the tachocline (which we assume at 0.7 solar radii) = $6.377 \cdot 10^8$ m, plus the distance between the surface and the tachocline (0.3 solar radii) which is $2.088 \cdot 10^8$. To derive the speed of the deep circulation we divide the sum by $t_{GS}$ minus 1 month (which we assume is the time for the flux to emerge from the tachocline to the surface at sunspot latitudes).

The calculated speed (solid line in Fig.5) is between 2.5 and 5.5 m/s, in good agreement with the speed estimated from the drift velocity of the sunspot bands toward the equator (Hathaway et al., 2003). Also shown in Fig.5 is the maximum sunspot number in the following sunspot cycle (dashed line). Vdeep shows a high positive correlation with the sunspot maximum following it – r=0.79 with p<0.001. This, according to Yeates et al. (2008) means that in the last 13 cycles diffusion has been more important than advection in the bottom part of the solar convection zone, or in other words, solar dynamo has been operating in diffusion-dominated regime near the tachocline.

We should note here that it seems this was not always the case. During the Maunder minimum, the correlation between the speed of the deep circulation and the amplitude of the following sunspot maximum was negative: a higher sunspot maximum followed after a slower deep circulation (Fig. 6).

To calculate the speed of the deep circulation during this period, the years of minima in 10Be during the Maunder minimum (Beer et al., 1998) were used as the years of maximum poloidal field, and the years of maxima in the group sunspot numbers (Hoyt and Schatten, 1998) as the years of maximum toroidal field (Georgieva and Kirov, 2007). The negative correlation means that in that period the solar dynamo operated in advection dominated regime which occurs when the speed of the deep circulation is relatively high and the diffusivity is relatively low, so increasing the speed means less time for the toroidal field to be generated in the tachocline (Yeates et al., 2007).

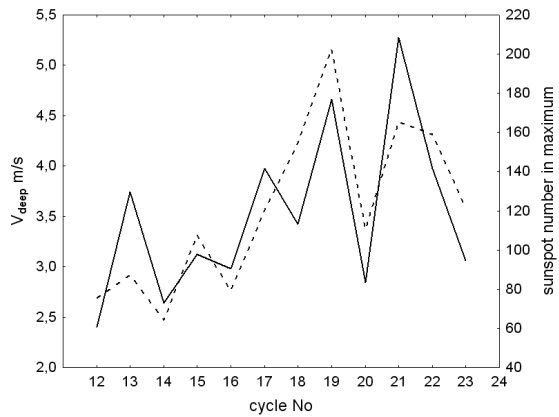
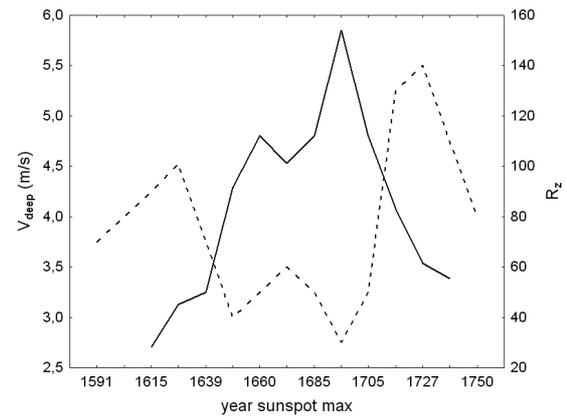

Fig.5: Amplitude of the sunspot maximum in consecutive sunspot cycles (broken line) and the speed of the deep meridional circulation preceding this sunspot maximum (solid line) in the case of advection dominated regime near the surface.

Fig.6: Speed of the deep meridional circulation (solid line) and amplitude of the sunspot maximum (broken line) during the Maunder minimum.

## B. Strongly diffusion-dominated regime near the surface

If the diffusivity is very high, the poloidal field diffuses directly to the tachocline, "short-circuiting" the meridional circulation. In this case the time for the field to diffuse through the convection zone is equal to $L^2/\eta_{surf}$ where L is the thickness of the convection zone ($0.3R_s$) and $\eta_{surf}$ is the diffusion coefficient in the bulk of the convection zone. Actually, the time from geomagnetic activity maximum on the sunspot declining phase and the following sunspot maximum, which we denote $t_{GS}$, includes not only the time for the poloidal field to diffuse from the surface to the base of the convection zone but also the time needed for the toroidal field to be generated from the poloidal field during its equatorward transport at the base of the convection zone, plus the time for the newly generated toroidal field to emerge from the tachocline to the surface. To get a rough estimate, we assume the speed of the deep poleward circulation as calculated for the case of advection dominated regime near the surface, and the distance the poloidal field travels at the surface from sunspot latitudes until it sinks to the base of the convection zone, and at the base of the convection zone until it is transformed into toroidal field both equal to 15° – between midlatitudes (30°) and the average highest latitudes of sunspot emergence (15°). Calculated in this way, the derived values of the average diffusivity in the bulk of the convection zone in the period 1868-2000 vary between $1.4 \times 10^8$ and $3.1 \times 10^8$ (the solid line in Fig.7). The ratio $\eta_{surf}/V_{surf}$ is between $1 \times 10^7$ and $6 \times 10^7$ m. This ratio is about two times higher than the value $\eta/u_0$ suggested by Hotta and Yokoyama (2010) to estimate the dominance of diffusion because $V_{surf}$ is the surface poleward speed averaged over latitude and over time, and is thus roughly half of $u_0$ which is the maximum surface poleward speed. The dashed line in Fig.7 presents the ratio $\eta_{surf}/u_0$ calculated with $u_0=2V_{surf}$. Most of the time both $\eta_{surf}$ and $\eta_{surf}/V_{surf}$ are below the respective values needed for strongly diffusion dominated regime near the surface, as estimated by Hotta and Yokoyama (2010).

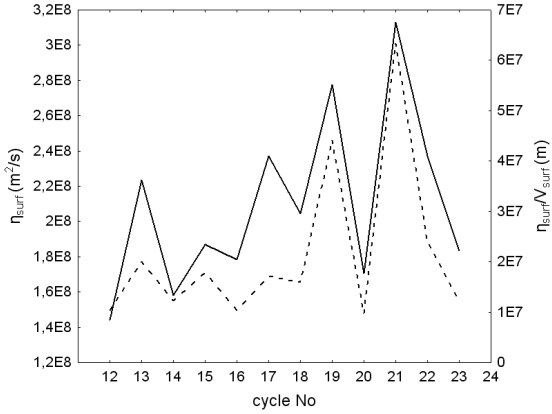

Fig.7. Average diffusivity in the bulk of the convection zone (solid line) and the ratio $\eta_{surf}/u_0$ (broken line) in case of strongly diffusion dominated regime near the surface.

## C. Moderately diffusion-dominated regime near the surface

In the case when the diffusivity in the upper part of the convection zone is high enough for a part of the poloidal field to short-circuit the meridional circulation, but still low enough so that another part of it does reach the poles, sinks there and is carried by the deep circulation equatorward, the sunspot maximum will be a superposition of the two surges of the toroidal field: generated from the poloidal field diffused across the convection zone, and from the poloidal field advected by the meridional circulation, occurring at times $t_{GS-D}$ and $t_{GS-A}$ after the geomagnetic activity peak on the sunspot declining phase, respectively. The diffusion time of the poloidal field across the convection zone is expected to be comparable to the advection time by the meridional circulation because the time scales of both turbulent diffusivity and meridional circulation arise from the same physics of turbulence in the convection zone (Jiang et al., 2008). Though close, $t_{GS-D}$ and $t_{GS-A}$ in general will not be exactly equal which will result in a double-peaked sunspot maximum as observed in many cycles. Gnevyshev (1967) suggested that all cycles have two peaks resulting from different driving mechanisms, and with varying time lag between them, so if this time lag is small the two peaks superpose and a single maximum is seen in the total global activity. But even in this case the two peaks are clearly visible when looking at different latitudinal bands. In the cycles studied by Gnevyshev (1967), the first peak appears almost simultaneously at all latitudes, while the second one, though higher, is only seen at low latitudes.

To distinguish between the two peaks, we use the Royal Greenwich Observatory - USAF/NOAA Sunspot Data compiled by NASA Marshall Space Flight Center (http://solarscience.msfc.nasa.gov/greenwch/bflydata.txt) and examine the evolution of sunspot activity in cycles 12-23 as expressed by the total area of sunspots in 25 latitudinal bins distributed uniformly in sine latitude, averaged over the two hemispheres. As shown e.g. by Temmer et al. (2006) and Norton and Gallagher (2010), the double peaked maximum is a real phenomenon that occurs in both hemispheres and is not due to a superposition of sunspot indices from the two, so we can use this averaging. On the other hand, Kane (2007) found that the peaks are often out of phase in the northern and southern hemispheres, so the averaging introduces some error, but it is acceptable for the purposes of the present study.

Fig. 8 is an illustration of the two peaks in sunspot area in cycle 16. One of the peaks (Fig.8a) centered around Carrington rotation 970 appears almost simultaneously in a latitudinal band from 26 to 18° and dissolves in 20-30 CR, while the other one first appears around CR 980 at 16° and until CR 1030 moves equatorward down to 4° with a maximum at 13° (Fig.8b). We attribute the peak appearing in a wide latitudinal band at time $t_{GS-D}$ after the geomagnetic activity peak to the toroidal field generated from the poloidal field diffusing at all latitudes during its transport to the poles at the surface, and the equatorward moving peak with maximum at time $t_{GS-A}$ after the geomagnetic activity peak to the toroidal field generated from the poloidal field advected all the way to the poles at the surface and equatorward at the base of the solar convection zone.

The picture presented in Fig.8 is observed in all solar cycles from 15 to 19: first activity in a wide latitudinal band at higher latitudes, followed by activity at lower latitudes moving further equatorward with time. In cycles 12-14 and after cycle 19 the order is reversed: first the advection-generated peak appears at higher latitudes moving equatorward, followed by the diffusion-generated peak in a wider band at lower latitudes. An example is presented in Fig. 9 – cycle 22. The sunspot area first peaks at 21° in CR 1806, then at 18° in CR 1811, at 16° in CR 1818, at 13° in CR 1833 (Fig.9a). The next peak occurs simultaneously around CR 1843 in a latitudinal band from 11 to 6°.

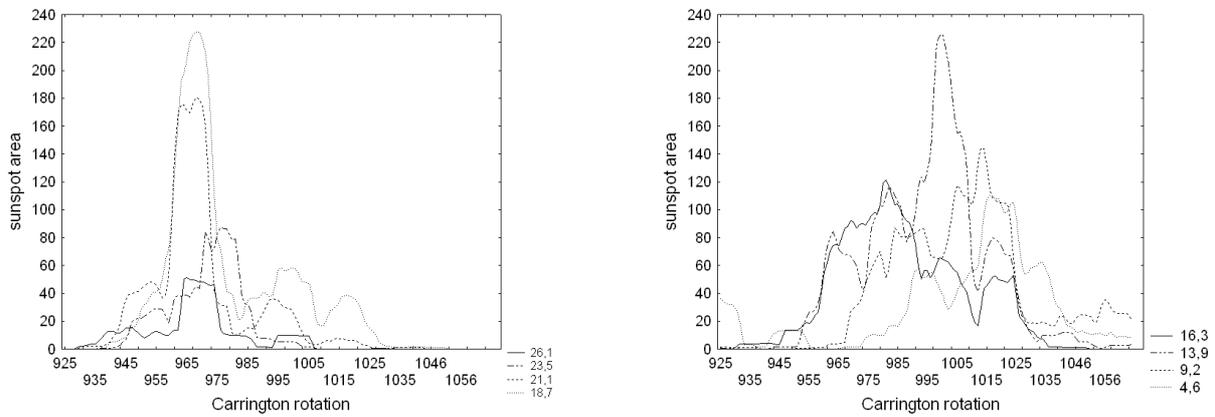

Fig.8. Two peaks in the sunspot area appearing at different latitudes (see the legend) in cycle 16: (a) diffusion–generated; (b) advection-generated.

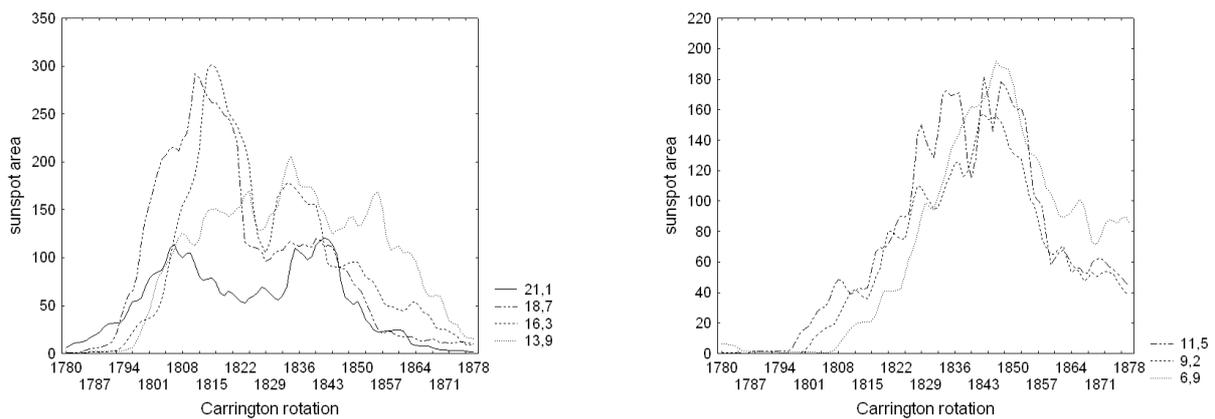

Fig.9. Two peaks in the sunspot area appearing at different latitudes (see the legend) in cycle 22: (a) advection-generated; (b) diffusion–generated.

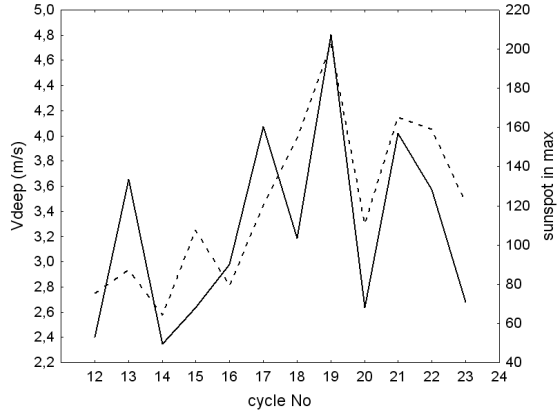

Fig.10. Amplitude of the sunspot maximum in consecutive sunspot cycles (broken line) and the speed of the deep meridional circulation preceding this sunsput maximum (solid line) in the case of moderately diffusion dominated regime near the surface.

From $t_{GS-D}$ and $t_{GS-A}$ we can estimate $\eta_{surf}$ - the average diffusivity in the upper part of the convection zone (the part involved in poleward motion), and $V_{deep}$ - the speed of the deep meridional circulation in the bottom part of the convection zone (the part involved in equatorward motion), respectively. We begin with $V_{deep}$ calculated in the same way as in the case of Advection-dominated regime near the surface described above, but with $t_{GS}$ replaced by $t_{GS-A}$. The results are presented in Fig.10 together with the maximum sunspot number in the following sunspot cycle.

Again, as when we suggested Advection dominated regime near the surface, $V_{deep}$ is positively correlated with the amplitude of the following sunspot cycle (r=0.75 with p=0.05) which confirms that in all this period solar dynamo has been operating in diffusion dominated regime near the tachocline.

To estimate the average diffusivity in the upper part of the convection zone, we apply the same approach as in the case of Strongly diffusion-dominated regime near the surface but with $t_{GS}$ replaced by $t_{GS-D}$, and with $V_{deep}$ calculated from $t_{GS-A}$: we assume that $t_{GS-D}$ - the time between the geomagnetic activity peak on the sunspot declining phase and the next diffusion generated sunspot peak - is equal to $L^2/\eta_{surf}$, the time it takes the poloidal field to diffuse to the base of the convection zone where L is the thickness of the convection zone (0.3$R_s$), plus the time needed for the toroidal field to be generated from the poloidal field during its equatorward transport at the base of the convection zone from 30° to 15° at speed $V_{deep}$, plus the time for the newly generated toroidal field to emerge from the tachocline to the surface which is again assumed to be one month.

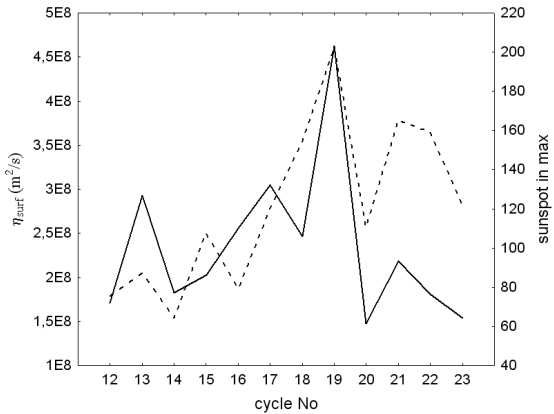

Fig.11. Average diffusivity in the bulk of the convection zone (solid line) and the maximum sunspot number in consecutive sunspot cycles (broken line) in case of moderately diffusion dominated regime near the surface.

The average diffusivity in the upper part of the convection zone $\eta_{surf}$ calculated in this way is presented in Fig. 11. It varies between 1.5 and 4.5x10$^8$ m$^2$/s and is again related with the amplitude of the sunspot maximum though the correlation is lower than when strongly diffusion dominated regime is assumed. An interesting feature observed in Fig.11 is that the relation between the diffusivity and the sunspot maximum changes after cycle 19: higher maximum sunspot number corresponds to lower diffusivity than up to cycle 19.

Now, when we have both the speed of the flow and the diffusivity in the upper part of the solar convection zone, we can evaluate the magnetic Reynolds number which is a measure of the relative importance of diffusion versus advection: $R_m = V_{surf} L / \eta_{surf}$ where L is the characteristic length scale. In this case L is the distance which the flow passes from maximum sunspot latitudes to the poles $\sim 9 \cdot 10^8$ m (Section 3.3.1). $V_{surf}$ varies between 5 and 20 m/s, and $\eta_{surf}$ between 1.5 and $4.5 \times 10^8$ m$^2$/s. Consequently, $R_m$ in the bulk of the solar convection zone is between 10 and 66.

Besides, if we have estimated the speed of the equatorward circulation in the lower part of the solar convection zone involved in equatorward meridional circulation, and if we know in what regime solar dynamo is operating near the tachocline, we could get some idea of the diffusivity around the tachocline. Diffusion dominated regime around the tachocline means $R_m < 1$, and with $R_m = V_{deep} L / \eta_{deep}$ this means $\eta_{deep} < V_{deep} L$. On the other hand, as mentioned above (see also Georgieva and Kirov, 2007), during the Maunder minimum solar dynamo was operating in advection dominated regime near the tachocline, which means $R_m > 1$ and $\eta_{deep} > V_{deep} L$. We can hypothesize that, in order for the dynamo to switch between the two regimes without drastic changes in its parameters, $R_m$ should be of order 1, that is, $\eta_{deep} \sim V_{deep} L$. With $V_{deep} \sim$ m/s and $L \sim 10^8$ m, this gives $\eta_{deep} \sim 10^8$ m$^2$/s. In other words, if our logic is correct, the diffusivity around the tachocline is the same order of magnitude as in the bulk of the convection zone, and the tachocline is probably a sharp boundary not only between the differentially rotating convection zone and the rigidly rotating radiative zone, but also between the high diffusivity in the convection zone and the low diffusivity in the radiative zone.

## 4. Long-term variations in solar activity

In Section 3 it was demonstrated that, as predicted by theory (Wang et al., 2002), the speed of the surface poleward meridional circulation is anticorrelated with the amplitude of the following sunspot maximum. Further, it was shown that in the last 12 sunspot cycles the speed of the deep equatorward circulation is positively correlated with the amplitude of the following sunspot maximum. Now we are looking at the sequence of relations in order to identify the factors ruling the way the solar dynamo operates.

**A. Surface meridional circulation → Polar field**

Fig.4 demonstrates that a higher sunspot maximum follows after a slower surface poleward circulation $V_{surf}$. The speed of the surface meridional circulation is important for the generation of the poloidal field from the toroidal field, so not only the sunspot maximum of cycle n+1 but also the polar field between sunspot maxima of cycles n and n+1 should be (negatively) correlated with the speed of the surface meridional circulation after the sunspot maximum of cycle n. This correlation should hold in any regime of the operation of the dynamo near the surface: In advection-dominated regime near the surface, all of the toroidal field of the next cycle is generated from the polar field advected by the surface meridional circulation to the poles, so the correlation between the surface circulation and the polar field mediates the correlation between the surface circulation and the next sunspot maximum. In diffusion dominated regime near the surface, both the polar field and the toroidal field of the next cycle will be produced from the same poloidal field which has been either advected by the surface meridional circulation (in the case of moderately diffusion dominated regime) or diffused (in the case of strongly diffusion dominated regime) to the poles. It seems that the number of polar faculae is not a reliable proxy for the polar field ( Jiang et al., 2008), but we already have 4 cycles of measured solar polar field (Wilcox Solar Observatory data obtained via the web site http://wso.stanford.edu courtesy of J.T. Hoeksema). Fig. 12 presents the relation between the surface poleward circulation and the maximum polar field following it. Though the statistics is small, clearly seen is the high negative correlation.

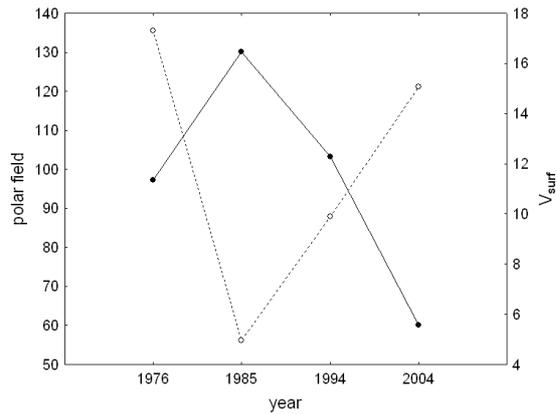

Fig.12. Strength of the polar field between the sunspot maxima of cycles n and n+1 (solid line) and the speed of the surface poleward meridional circulation after the maximum of cycle n (broken line).

### B. Polar field → Deep meridional circulation

According to the so-called Malkus-Proctor mechanism (Malkus and Proctor, 1975) the back reaction of the Lorenz force generated by the magnetic field on the velocity field should lead to correlation between the magnitude of the solar polar field and the speed of the following deep meridional circulation (Charnonneau, 2005). We have only 3 cycles with measured polar fields and the speed of the deep equatorward meridional circulation following it, and they show a very high correlation (Fig.13). Of course, 3 points are not enough to draw a conclusion, but the expected correlation between the surface poleward circulation and the following polar field should also mean a correlation between the surface poleward circulation and the following deep equatorward circulation. Indeed, as demonstrated in Fig.14, the two are anticorrelated (r=0.7 with p=0.005) which also confirms the correlation between the polar field and the deep equatorward circulation after its maximum.

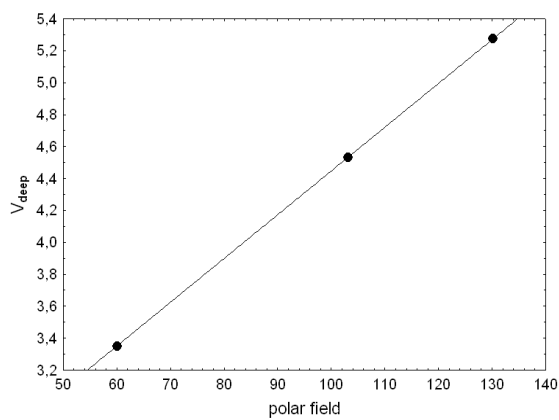

Fig.13. Dependence of the speed of the deep meridional circulation on the strength of the polar field.

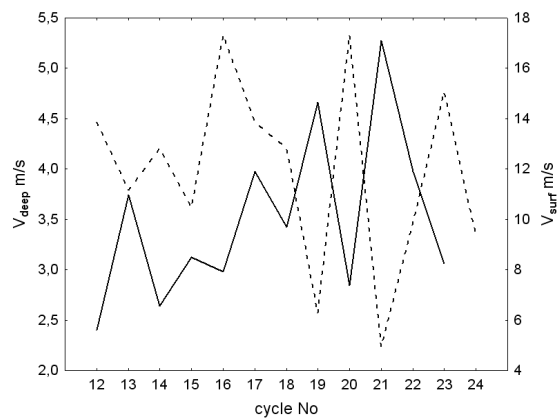

Fig.14. Speed of the surface meridional circulation between sunspot maximum and the following geomagnetic activity maximum (solid line) and the speed of the deep meridional circulation between this geomagnetic activity maximum and the next sunspot maximum.

## C. Deep meridional circulation → Next sunspot maximum

As demonstrated in Fig.5 and Fig.10, the amplitude of the sunspot cycle is highly correlated with the speed of the deep equatorward circulation preceding it which confirms that in all this period solar dynamo has been operating in diffusion dominated regime near the tachocline. However, as evident from Fig.6, during the Maunder minimum (and during other grand minima as well, as far as we can judge from geomagnetic activity reconstructions (Georgieva et al., 2009)), the amplitude of the sunspot cycle was anticorrelated with the speed of the deep equatorward circulation which indicates that then solar dynamo at the tachocline was operating in advection dominated regime. During the Maunder minimum the diffusivity in the bulk of the convection zone, estimated in the same way as for the recent cycles, was ~$2.10^8$ m$^2$/s, not different from the recent cycles. What was different was the very slow surface circulation, comparative with the speed of the deep circulation, so that during this period the ratio $V_{surf}/V_{deep}$ was around 1 (Fig.15).

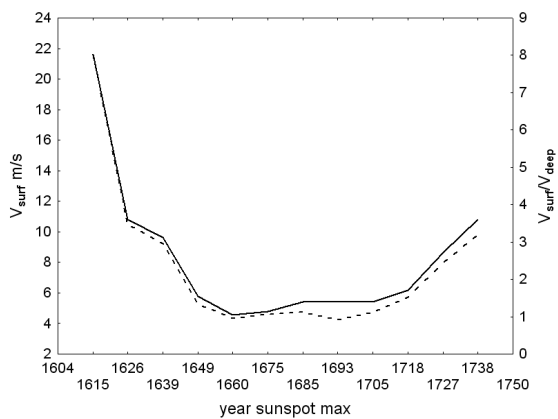

Fig.15. Speed of the surface meridional circulation (solid line) and ratio of the surface to deep circulation speeds (broken line) during the Maunder minimum.

The end of the Maunder minimum was marked by a gradual increase in the speed of the surface circulation and by the recovery of the positive correlation between Vdeep and the amplitude of the sunspot maximum (Fig.6). At present it is not clear what caused the drop in the speed of the surface circulation, but it seems that this drop was the factor which initiated the Maunder minimum.

## D. Sunspot maximum → Surface meridional circulation

There is no correlation at all between the amplitude of the sunspot maximum and the speed of the surface meridional circulation after it (Fig.16).

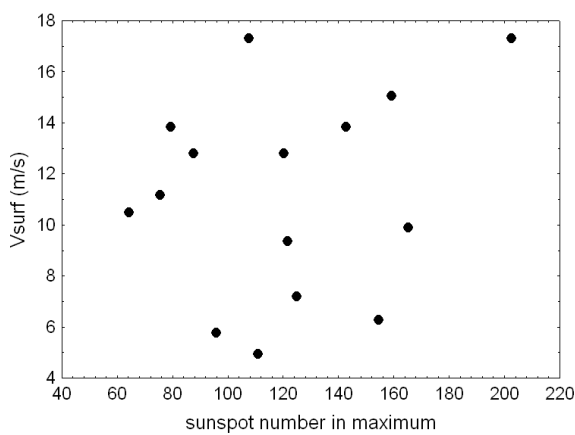

Fig.16. Dependence of the speed of the surface meridional circulation on the amplitude of the sunspot maximum preceding it.

This finding is in agreement with the lack of correlation found by Jiang et al. (2008) between the maximum sunspot number of a cycle and the polar field at the end of this cycle whose strength is determined by the surface circulation after the sunspot maximum.

The lack or correlation between the sunspot maximum and the speed of the surface meridional circulation after it limits the sunspot cycle "memory" – the correlations between the parameters of cycle n and the amplitude of cycles n+1, n+2, etc. The chain of correlations is as follows: Surface meridiuonal circulation → Polar field → Deep meridional circulation → Next Sunspot Maximum, and there the chain breaks. Therefore, the sunspot cycle has a "short memory".

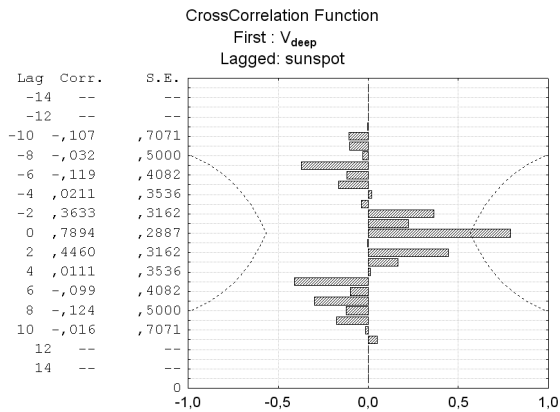

This is confirmed by the cross-correlation function of the deep meridional circulation and the maximum sunspot number: the sunspot cycle maximum is correlated to the speed of the deep circulation preceding it, with no contribution from earlier cycles (Fig.17).

Fig.17. Cross-correlation function of the speed of the deep meridional circulation and the amplitude of the sunspot maximum following it. Also shown are the 95% confidence limits.

**E. Advection-generated and diffusion-generated sunspot activity**

As commented in Section 3.3.2.C and demonstrated in Fig.8 and 9, in moderately diffusion-dominated regime near the surface the sunspot maximum is a result of the superposition of two surges of toroidal field: one generated from poloidal field advected all the way to the poles, down to the tachocline and back to sunspot latitudes, and the other one from poloidal field duffused directly toward the tachocline at midlatitudes. According to Gnevyshev (1967), the first maximum (which we identify as diffusion-generated) always appears earlier and at higher latitudes, while the second one (which we identify as advection-generated) appears later, at lower latitudes and as a rule is stronger than the first one. Our examination of cycles 12-23 shows that this is not always the case. The diffusion-generated maximum appears at higher latitudes and remains there, while the advection-generated maximum appears at lower latitudes and moves further equatorward during the secular increase of solar activity (from cycle 15 to cycle 19). In periods of solar activity decrease, the advection-generated maximum appears at higher latitudes and moves equatorward but without reaching any close to the equator, while the diffusion-generated maximum appears at lower latitudes and remains there, but again without reaching the equator. We can speculate that the latter case is an additional factor for decreasing solar activity as fewer leading polarity sunspots appear close to the equator and have the chance to be canceled with their counterparts of the opposite hemisphere. Obviously the different behavior of the diffusion-generated and advection-generated toroidal field is due to variations in the way solar dynamo operates, but at present we cannot find a good explanation.

**4. Discussion**

We have demonstrated the results of a method for deriving of the long-term variations in the solar surface and deep meridional circulation, and the diffusivity in the bulk of the solar convection zone, based on the varying time-lag between sunspot maximum and geomagnetic activity maximum. The derived speed of the surface meridional circulation agrees with the direct measurements, and the speed of the deep meridional circulation agrees with estimations based on drift velocity of the sunspot bands toward the equator. This gives us confidence in the reliability of our method for estimating the long-term variations in solar meridional circulation. The diffusivity in the bulk of the convection zone which we have estimated is in agreement with the calculations of Ruzmaikin and Molchanov (1997) based on granulation and supergranulation, and with the number of arguments in favor of this order of magnitude presented in Jiang et al. (2007). A somewhat unexpected result is that the tachocline is a sharp boundary also for diffusivity, not only for rotation, but this is not improbable taking into account the very different physical parameters in the radiative and convection zones.

The magnetic Reynolds number calculated from the derived values of the surface velocity and the diffusivity in the bulk of the convection zone is of order 10, so the regime of operation there is diffusion dominated, but not strongly diffusion dominated. An argument against the strongly diffusion dominated regime near the surface is that diffusion alone cannot explain the observed strongly peaked polar fields around sunspot minimum (De Vore et al., 1984). Finally, a moderately diffusion dominated regime in which toroidal field is generated by both poloidal field diffused at midlatitudes directly toward the tachocline, and poloidal field advected to the pole at the surface, down to the tachocline and equatorward at the base of the convection zone, is the only natural explanation of the double peaked sunspot cycle - the so called Gnevyshev gap.

The comparison of the long-term variations in sunspot activity and in the derived speeds of the surface and deep solar meridional circulation makes it possible to determine the sequence of processes ruling the amplitude of the sunspot cycle. We find that the main factor is the speed of the surface poleward circulation which is related to the strength of the polar field, which modulates the speed of the deep equatorward circulation, which determines the magnitude of the toroidal field generated from the poloidal field during its transport at the base of the convection zone. In turn, the speed of the surface poleward circulation is not at all correlated with the magnitude of the preceding sunspot maximum, so the solar dynamo has a one cycle only long memory. The role of the surface meridional circulation in modulating the sunspot cycle is accounted for in some models (e.g. Charbonneau and Dikpati, 2000), so stochastic fluctuations are introduced in Vsurf to represent its variability. However, the quasi-periodic changes in Vsurf imply that there may be some physical factor modulating Vsurf which is not yet identified with confidence. Not clear either is what determines the variations in the temporal and spatial evolution of the advection-generated and diffusion-generated toroidal field, but it seems that they are related to solar dynamics and in turn contribute to the modulation of the sunspot cycle.

The speed of the surface meridional circulation seems to determine also the regime of operation of the solar dynamo at the base of the convection zone. The regime – diffusion dominated or advection dominated – depends on the relative importance of diffusion versus advection, and can be identified based on the correlation between the speed of the deep meridional circulation and the amplitude of the following sunspot maximum. We find that this correlation is positive – a faster deep circulation is followed by a higher sunspot maximum, therefore at least in the last 12 cycles solar dynamo has been operating in diffusion dominated regime. However, this has not always been so: during the Maunder minimum, and probably also during other grand minima, dynamo was operating in advection dominated regime characterized by negative correlation between the speed of the deep meridional circulation and the amplitude of the following sunspot maximum. According to the chain of relations outlined above, the speed of the deep meridional circulation depends on the speed of the surface meridional circulation. In all studied periods the two are anticorrelated, but their ratio is not constant. During the Maunder minimum the speed of the surface circulation dropped to extremely low values and the ratio $V_{surf}/V_{deep}$ fell to around 1. It is not clear whether this is due to stochastic fluctuations or to the action of some physical factor. This is for now an open question which requires more investigation, especially because it is related to the eventual possibility for long-term forecast of solar activity.

**Acknowledgements**

The authors are grateful to the referees for the very helpful comments which have led not only to the improvement of the original paper but also to new ideas included in the final version.

**References**

Babcock, H.W., 1961. The topology of the sun's magnetic field and the 22-year cycle, The Astrophysical Journal 133, 572-587.

Basu, S., Antia, H. M., 2003. Changes in Solar Dynamics from 1995 to 2002. The Astrophysical Journal, Volume 585, Issue 1, pp. 553-565.

Beer, J., Tobias, S., Weiss, N., 1998. An Active Sun Throughout the Maunder Minimum. Solar Physics 181 (1), 237-249.

Charbonneau, P. 2005. Dynamo Models of the Solar Cycle. Living Reviews im Solar Physics 2, (2005), 2. Online Article: cited 0n 30 July 2009. http://www.livingreviews.org/lrsp-2005-2

Charbonneau, P., Dikpati, M., 2000. Stochastic Fluctuations in a Babcock-Leighton Model of the Solar Cycle. The Astrophysical Journal 543 (2), 1027-1043.

Cliver, E.W., 1994: Solar Activity and Geomagnetic Storms: The First 40 Years. Eos, Transactions, AGU, 75 (49), 569, 574-575.

Echer, E., Gonzalez, W. D., Gonzalez, A. L. C., Prestes, A., Vieira, L. E. A., Dal Lago, A. Guarnieri, F. L., Schuch, N. J., 2004. Long-term correlation between solar and geomagnetic activity. Journal of Atmospheric and Solar-Terrestrial Physics 66 (12), 1019-1025.

Feynman, J., 1982. Geomagnetic and solar wind cycles, 1900-1975. Journal of Geophysical Research 87, 6153-6162.

Georgieva, K., Kirov, B., Obridko, V.N., Shelting, B.D., 2009. What can we learn about solar dynamo and solar influences on climate from geomagnetic data. All-Russian annual Conference "Solar and Solar-Terrestrial Physics - 2008", 53-56.

Georgieva, K., Kirov, B., 2006. Solar Activity and Global Warming Revisited. Sun and Geosphere 1 (1), 12-26.

Georgieva, K., Kirov, B., 2007. Long-term variations in solar meridional circulation from geomagnetic data: implications for solar dynamo theory. arXiv:physics/0703187v2 [physics.space-ph]

Gleissberg, W., 1944. A table of secular variations of the solar cycle, Terrestrial Magnetisn and Atmospheric .Electricity 49, 243-244.

Gnevyshev, M. N., 1967. On the 11-Years Cycle of Solar Activity, Solar Physics 1 (1), 107-120.

González Hernández, I., Komm, R., Hill, F., Howe, R., Corbard, T., Haber, D. A., 2006. Meridional Circulation Variability from Large-Aperture Ring-Diagram Analysis of Global Oscillation Network Group and Michelson Doppler Imager Data. The Astrophysical Journal 638, 576-583.

Hathaway, D., 1996. Doppler measurements of the Sun's meridional flow. The Astrophysical Journal 460, 1027-1033.

Hathaway, D., Nandy, D., Wilson R., Reichmann, E., 2003. Evidence that a deep meridional flow sets the sunspot cycle period, The Astrophysical Journal 589, 665-670.

Hathaway, D., Nandy, D., Wilson, R., Reichmann, E., 2004. Erratum: "Evidence that a Deep Meridional Flow Sets the Sunspot Cycle Period", The Astrophysical Journal 602 (1), 543-543.

Hotta, H., Yokoyama, T., 2010. Importance of Surface Turbulent Diffusivity in the Solar Flux-Transport Dynamo. The Astrophysical Journal 709 (2), 1009-1017.

Hoyt, D. V., Schatten, K. H., 1998. Group Sunspot Numbers: A New Solar Activity Reconstruction. Solar Physics 181 (2), 491-512.

Ivanov, E.V., Obridko, V.N. and Shelting, B.D., 2002. Meridional drifts of large-scale solar magnetic fields and meridional circulation, Proc. 10th European Solar Physics Meeting 'Solar Variability: from Core to Outer Frontiers", Prague, Czech Republic, 9-14 September 2002 (ESA SP-506, December 2002), 851-854.

Javaraiah, J., 2010. Long-term variations in the mean meridional motion of the sunspot groups. Astronomy and Astrophysics 509, id.A30.

Javaraiah, J., Ulrich, R. K., 2006. Solar-Cycle-Related Variations in the Solar Differential Rotation and Meridional Flow: A Comparison, Solar Physics 237 (2), 245-265.

Jiang, J., Chatterjee, P., Choudhuri, A.R., 2008. Solar activity forecast with a dynamo model. Monthly Notices of the Royal Astronomical Society 381 (4), 1527-1542.

Kane, R.P., 2007. Latitude Dependence of the ``Gnevyshev'' Peaks and Gaps. Solar Physics 245 (2), 415-421.

Kishcha, P. V., Dmitrieva, I. V., Obridko, V. N., 1999. Journal of Atmospheric and Solar-Terrestrial Physics 61 (11), 799-808.

Leighton, R., 1969. A Magneto-Kinematic Model of the Solar Cycle. The Astrophysical Journal 156, 1-26.

Makarov, V. I., Tlatov, A. G., Sivaraman, K. R., 2001. Does the Poleward Migration Rate of the Magnetic Fields Depend on the Strength of the Solar Cycle?, Solar Physics 202 (1), 11-26.

Malkus, W.V.R., Proctor, M.R.E., 1075. The macrodynamics of α-effect dynamos in rotating fluids. Journal of Fluid Mechanics 67 (3), 417-443.

Mayaud, P.N., 1972. The aa indices: A 100 years series characterizing the magnetic activity. Journal of Geophysical Research 77, 6870–6874.

Miesch, M.S., Elliott, J.R., Toomre, J., Clune, T.L., Glatzmaier, G.A., Gilman, P.A., 2000. Three-dimensional Spherical Simulations of Solar Convection. I. Differential Rotation and Pattern Evolution Achieved with Laminar and Turbulent States. The Astrophysical Journal 532, 593-615.

Norton, A. A., Gallagher, J. C., 2010. Solar-Cycle Characteristics Examined in Separate Hemispheres: Phase, Gnevyshev Gap, and Length of Minimum, Solar Physics 261 (1), 193-207.

Parker, E.N., 1955. Hydromagnetic dynamo models, The Astrophysical Journal 122, 293–314.

Richardson, I. G., Cane, H. V, W. 2002. Sources of geomagnetic activity during nearly three solar cycles (1972-2000). Journal of Geophysical Research (Space Physics) 107 (A8), SSH 8-1, DOI 10.1029/2001JA000504.

Ruzmaikin, A., Molchanov, S.A., 1997. A model of diffusion produced by a cellular surface flow. Solar Physics 173, 223–231.

Ruzmaikin, A., Feynman J., 2001. Strength and phase of the solar dynamo during the last 12 cycles. Journal of Geophysical Research 106 (A8), 15,783–15,789.

Švanda, M., Kosovichev, A.G., Zhao, J., 2007. Speed of Meridional Flows and Magnetic Flux Transport on the Sun, The Astrophysical Journal 670 (1), L69-L72.

Temmer, M., Rybák, J., Bendík, P., Veronig, A., Vogler, F., Pötzi, W., Otruba, W., Hanslmeier, A., 2006. Hemispheric Sunspot Numbers 1945--2004: data merging from two observatories. Central European Astrophysical Bulletin 30, 65-73.

Wang, Y.-M., Sheeley, N. R., Jr., 1990. Magnetic flux transport and the sunspot-cycle evolution of coronal holes and their wind streams. Astrophysical Journal, Part 1 365, 372-386.

Wang, Y.-M., Sheeley, N.R., Nash, A.G., 1991. A new solar cycle model including meridional circulation, The Astrophysical Journal 383, 431-442.

Wang, Y.-M., Sheeley, N. R. Jr., Lean, J., 2002. Meridional flow and the solar cycle variation of the Sun's open magnetic flux. The Astrophysical Journal 580, 1188–1196.

Webb, D. F., 2002. CMEs and the solar cycle variation in their geoeffectiveness. In: Proceedings of the SOHO 11 Symposium on From Solar Min to Max: Half a Solar Cycle with SOHO, 11-15 March 2002, Davos, Switzerland. A symposium dedicated to Roger M. Bonnet. Edited by A. Wilson, ESA SP-508, Noordwijk: ESA Publications Division, 409 – 419.

Yeates, A. R., Nandy, D., Mackay, D. H., 2008. Exploring the Physical Basis of Solar Cycle Predictions: Flux Transport Dynamics and Persistence of Memory in Advection versus Diffusion Dominated Solar Convection Zones, The Astrophysical Journal, 673 (1), 544-556.

Zhao, J. and Kosovichev, A., 2004. Torsional oscillation, meridional flows, and vorticity inferred in the upper convection zone of the Sun by time-distance helioseismology, The Astrophysical Journal 603, 776-784